\documentstyle[11pt,aaspp4,psfig]{article}  

\begin{document}
 
\title{EUVE Investigation of Three Short-Period Binary Stars}

\author{\sc Slavek M. Rucinski}
\affil{Eureka Scientific Inc.\\
2452 Delmer St., Oakland, CA 94602--3017}

\centerline{\today}

\begin{abstract}
EUVE satellite spectroscopic  observations (SW, MW and LW bands
covering 80 -- 160, 170 -- 350 and 450 \AA\  with resolutions 
0.5, 1 and 2 \AA) and photometric observations 
(Deep Sky Survey, broad-band 70 -- 140 \AA) have been 
obtained for two contact, 44i~Boo~B and VW~Cep, 
and one detached, ER~Vul, close binary stars. 
All three systems have orbital periods shorter 
than one day and thus are expected to show ``saturated'' levels of 
chromospheric, transition-region and lower-corona 
emissions. The spectroscopic data were of 
sufficient quality for an attempt at an emission-measure 
determination only for 44i~Boo~B. This 
determination, based entirely on iron lines, 
and utilizing Singular Value Decomposition formalism 
developed by Schmitt et al. (1996) indicates lack of any dominating 
temperature regime with the emission measure rising from $\log T \simeq 6$
to $\log T \simeq 7.2$. However, strong dependence of the resulting
emission-measure curve on the inclusion of individual
lines formed at high temperatures 
casts some doubts about the quality of the solution.
A comparison of the line strengths have been made for 
selected strongest chromospheric, transition-region and lower-coronal
emission lines; it included the data for the 
single, rapidly-rotating star AB~Dor which is the only such star with
the rotation period shorter than one day which has been
observed with the EUVE.
If the single-epoch observations are representative, 
the results indicate that AB~Dor is under-active relative to 
the three binary stars which show similar levels of activity.
\end{abstract}


\section{INTRODUCTION: VERY CLOSE BINARY STARS }
\label{into}

Very close, synchronized, late spectral-type binaries 
offer us laboratories of stars that are forced to 
rotate much more rapidly than would be normally 
encountered among late-type stars. Except for 
very young stars which still contain large amounts 
of original angular momentum from recent formation, 
old stars of the F, G 
and K spectral types tend -- on the average -- to rotate progressively 
less rapidly with the advancing spectral type due 
to increased efficiency of the magnetic-wind braking 
for lower effective temperatures. Contact 
binary systems are a most interesting exception in 
this respect as they obey the orbital period -- 
color relation which is just opposite to this trend, 
with the cooler stars rotating more rapidly. This 
unusual combination leads to a very large range in 
observed levels of magnetic activity among contact 
binaries, with highest levels well in the saturated 
regime (Vilhu \& Walter \markcite{vw87}1987). Two such systems are the
subject of this paper. They are compared with a very close, but
detached binary and with a single, rapidly-rotating young star. All
four stars have rotation periods shorter than one day.

There have been several reviews about contact binaries 
(Rucinski \markcite{ruc85a} 
\markcite{ruc93}1985a, 1993) and about some aspects of 
their magnetic activity (Rucinski \markcite{ruc85b}1985b). 
At this point it is sufficient to stress that, although these 
are binary stars, they should be considered as single 
entities from the point of view of their internal structure because of
the on-going (but hidden to observers) processes of mass and energy
exchange between the components. These
internal processes of energy and mass transfer 
within such structures are poorly understood, but this seems 
to be of relatively minor importance for 
characterization of activity of contact binaries which tends to follow
the basic 
trends established for non-contact stars. It is of relevance that 
periods and colors of contact systems correlate 
as expected for components that are not too 
distant from the Main Sequence: the small, short-period 
systems have low temperatures whereas the long-period systems 
are relatively hot and bright. The 
sharp, currently unexplained, short-period cutoff 
in the period distribution is observed  at 0.22 day corresponding to
spectral type K5. Most of the contact systems have orbital periods
concentrated within $0.25 < P < 0.5$ day where the spectral types
range between middle K and late A, with few reaching periods of
one day where the spectral types are middle A. 
Such contact systems are called the W~UMa-type 
binaries, in contrast to the 
continuation of this sequence to early-type contact 
binaries, with periods of a few days and spectral 
types as early as O-type. The latter are of no interest 
in the present context, so that in this paper the 
names of W~UMa-type and contact binaries will be used interchangeably. 

They new results based on the statistics for a 
volume-limited sample of  the contact systems in the 
OGLE micro-lensing project (Rucinski 
\markcite{ruc97a}\markcite{97b}1997a, 1997b) suggest 
that they are about 3 times more common in space 
than previously thought. The previous numbers 
based on the sky-field sample suggested the apparent 
density of one W~UMa system per 1000 
normal dwarfs (Duerbeck  \markcite{duer84}1984). 
The new results imply that the statistics of 
bright systems is skewed by difficulties of detection 
of low-amplitude systems,  which remain to be 
discovered in the all sky sample. This sample 
seems to be complete to relatively bright level of 
about $V \simeq 8$, where the apparent frequency is indeed about one
W~UMa system per 300 late-type dwarfs, but 
shows selection effects for fainter stars.

The above remarks have only a parenthetical significance, 
as the EUVE sky is limited to the 
nearest and usually brightest 
objects. In addition, from the large spectral 
range observed for the contact binaries, 
appreciable chromospheric and coronal activity is 
expected only for the coolest systems. The 
present paper discusses  the EUVE results for two 
late-type contact binary systems,  44i~Bootis~B 
(P = 0.268 day) and VW~Cephei (P = 0.278 day), 
which had been detected in the EUVE surveys 
and were judged bright enough for spectral observations. 
Observations of the third nearby system, 
$\epsilon$~Coronae Austrinae (HR~7152) with 
the spectral type F and orbital period 0.591 day, were 
also proposed, but were not granted EUVE time, 
as the star had not been detected in the surveys. It 
is not clear whether this non-detection was due to 
the early spectral type of the star or to its sky position at 
$(l,b) = (0^\circ,\,-17^\circ)$, that is toward the ``wall'' of 
local absorption in the solar vicinity (Warwick et al. 
\markcite{war93}1993). The EUVE observations of the 
two contact binaries are compared here
with observations of a close, but 
detached binary ER Vulpeculae (P = 0.698 day). 
This system is frequently wrongly classified 
as a contact binary, but there is no question that both 
components are well within their respective Roche lobes 
(Hill et al. \markcite{hill90}1990). We note 
that for all three systems, the orbital periods 
can be identified with the rotation periods due to sychronization
by the tidal forces. In the case 
of ER Vul, this implies a rotation rate some 40 times 
faster than solar. For the contact binaries, the 
rotation is still more rapid reaching some 
100-times solar for VW~Cep and 44i~Boo~B. 
The EUVE observations 
of the latter star which are described here
have already been discussed and 
interpreted within a context of other active 
binaries by Dupree \markcite{dup96a} \markcite{dup96b}(1996a, 1996b). 

The paper is organized as follows:
Section~\ref{description} describes the objects and the observations,
and Section~\ref{dss} discusses the EUV (70 -- 140 \AA, DSS) light curves.
A discussion of the neutral-hydrogen column densities is
presented in Section~\ref{hydrogen}. The main discussion of the EUV
spectra and the spectroscopic results for the three binary stars 
is in Section~\ref{spectra}. The next Section~\ref{comparison}
 compares the line-intensity results for the three binary 
systems and for the rapidly-rotating single star, AB~Dor whose EUVE
observations were described and analyzed before (Rucinski et
al.\markcite{abdor} 1995). The last
Section~\ref{conclusions} contains conclusions and summary of the
paper.

\section{EUVE OBSERVATIONS}
\label{description}

All the spectral and the Deep Sky Survey data 
have been processed automatically by the Center for 
EUV Astrophysics pipeline software to the level 
Quick Position Oriented Event (QPOE) files. The 
data have been manually extracted from the QPOE files 
during a visit to the Center for Extreme-Ultraviolet 
Astrophysics, Berkeley, California in February 1996. 
The software version used was \#1.6 and the calibration 
dataset was \#1.11.The reasons for re-processing 
was the usual issue, for EUVE observations,
 of the optimum placement of the 
extraction slit, with the added improvement in the 
Earth blockage temporal filter. 
The latter was particularly important for the case of 
VW~Cep, as the automatic software rejected a 
large fraction of data which were obtained at low view angles. 
The blockage zenith angle was changed 
uniformly for all data from $98^\circ$ to $107^\circ$.

Several details specific for the program stars 
as well as for the EUVE observations are listed in 
Table~\ref{tab1}. Since all these stars show period variations, 
observations of the moments of eclipse 
centers were collected from the recent literature 
and initial zero phases and periods were adopted 
for each star, as in Table~\ref{tab1}. This was particularly 
important for 44i~Boo~B which, in addition to apparently 
random period changes, shows a systematic 
period variation due to the barycentric motion in a 
visual binary system. For each system, some effort was 
made to find as recent predictions of the 
moments of  deeper eclipses as possible. The zero 
phase was then chosen to correspond to moment 
just before the start of the EUVE observations so 
that the calculated phases could be conveniently 
used as a measure of time within each EUVE run.

The stellar data have been taken from the papers 
cited below. For estimates of the bolometric 
fluxes, the formula: $f_{bol} = 2.5 \times 10^{-0.4 m_{bol}}$ 
was used, with $m_{bol} = V_{max} + B.C.$, where the 
bolometric corrections were from the most recent discussion of 
Flower \markcite{phil96}(1996). Some additional 
discussion of distances and maximum 
magnitudes is in Rucinski \markcite{ruc94}(1994).

A few additional details about the three systems and information about
timing of their orbital motions are given below:

\begin{description}
\item[44i~Boo~B:] The star is the fainter member of 
the visual binary system ADS~9494. Because 
of the proximity of the brighter companion 
of spectral type about G1V, the color and spectral type 
of the contact binary are uncertain, but 
must be very late due to shortness of the orbital period. 
Normally, it is assumed that the contact 
binary produces all activity-related phenomena, and that the visual
companion A does not contribute anything; the 
same assumption has been made in this paper. The stellar 
data as well as the distance have been adopted 
following the extensive discussion by Hill, 
Fisher \& Holmgren \markcite{hill89a} (1989a). $B-V 
\simeq 0.7$ assumed by Rucinski 
\markcite{ruc94}(1994) was probably too blue, and the present 
entry is still very uncertain. The zero-phase 
for eclipses was adopted using observations by 
Rovithis-Livaniou et al. \markcite{rl95}(1995), 
which was obtained only a few days after the EUVE 
observations. The period is also from a 
recent discussion by Oprescu et al. \markcite{op96}(1996). 

\item [VW~Cep:] There is a faint companion in 
the system which has been assumed not to contribute 
to the activity levels observed in the EUVE 
band. The stellar and distance data are after Hill 
\markcite{hill89b}(1989). The eclipse 
time prediction  was taken from Aluigi et al. 
\markcite{al94} (1994)  which cites 
several other recent determinations. This prediction  agrees 
well with continuing observations of 
VW~Cep just before and after the EUVE run by Kiss et al. 
\markcite{kiss95}(1995) and Oprescu et al. \markcite{op96}(1996).

\item [ER Vul:] The system has been extensively 
studied by Hill, Fisher \& Holmgren 
\markcite{hill90} (1990). The primary-eclipse 
ephemeride from this paper has been checked 
against the new determination of Zeinali et al. 
\markcite{zei95}(1995) and found still applicable. 
\end{description}

\section{DSS LIGHT CURVES}
\label{dss}

The complete description of the EUVE 
instruments is in Bowyer \& Malina 
\markcite{bm91}(1991). The Deep Sky Survey 
(DSS) channel providing photometric information 
in the band 70 -- 140 \AA\   operated
during the spectral observations. It made possible
monitoring of the stars for possible 
flaring activity and for any orbital-phase dependent modulation 
which could be due to concentration of 
active regions in some localized areas. The DSS data have 
been binned into 100 second bins. Partial 
bins shorter than 50 seconds have been discarded to 
ensure reasonable total counts per bin.
For all three stars the orbital phases were used as the 
time coordinates; the relevant data are given in Table~\ref{tab1}.

\begin{figure}      
\centerline{\psfig{figure=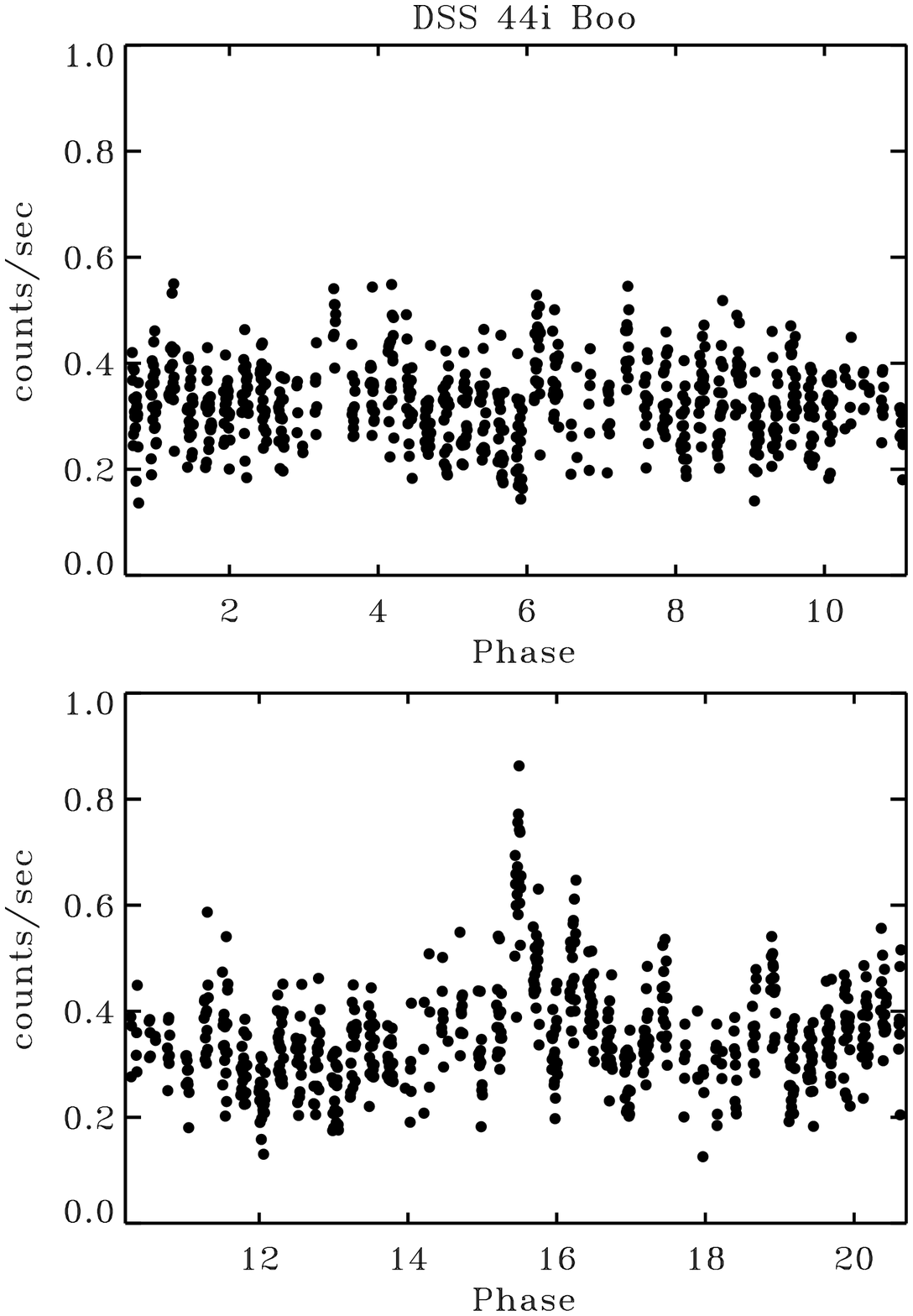,height=6.0in}}
\vskip 0.5in
\caption{\label{fig1}
The Deep Sky Survey (DSS, 70 -- 140 \AA) light curve of 44i~Boo~B. The
time scale is in orbital periods counted from the initial epoch, as in
Table~\ref{tab1}. The data have been binned into 100 second intervals
and the bins containing data for more 
than 50 seconds were used to derive the count rates.}
\end{figure}

\begin{figure}      
\centerline{\psfig{figure=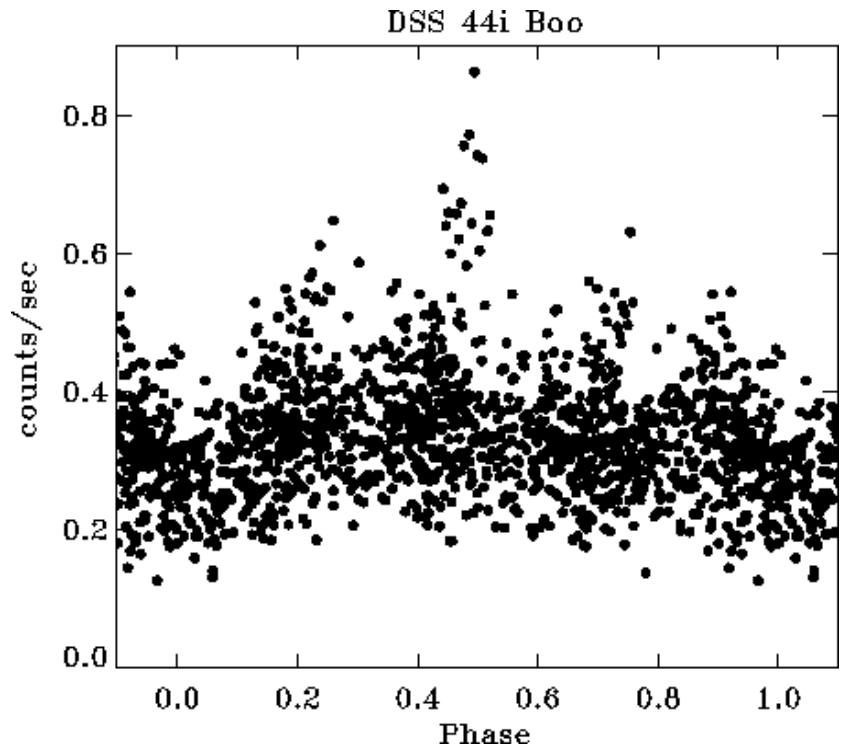,height=4.0in}}
\vskip 0.5in
\caption{\label{fig2}
The same as in Fig.~\ref{fig1}, but with time folded into one orbital
period. Note that clumping of the data is due to the commensurability
of the binary and satellite periods. However, the tendency for stronger
activity at phases around 0.5 seems to be real.}
\end{figure}

\begin{figure}      
\centerline{\psfig{figure=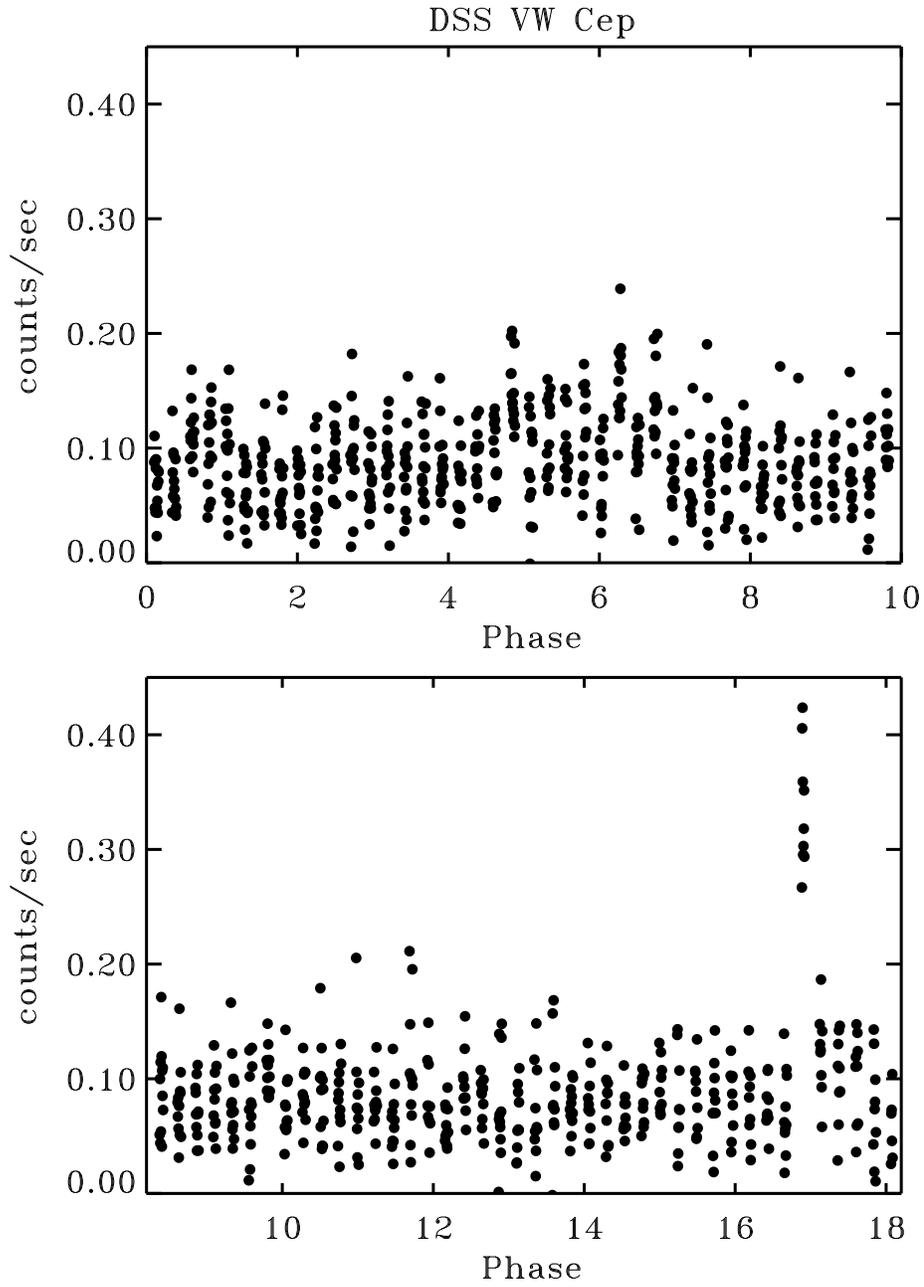,height=6.0in}}
\vskip 0.5in
\caption{\label{fig3}
The DSS light curve for VW~Cep. The system brightened
briefly in EUV close to phase the primary eclipse of cycle number 17.}
\end{figure}

\begin{figure}      
\centerline{\psfig{figure=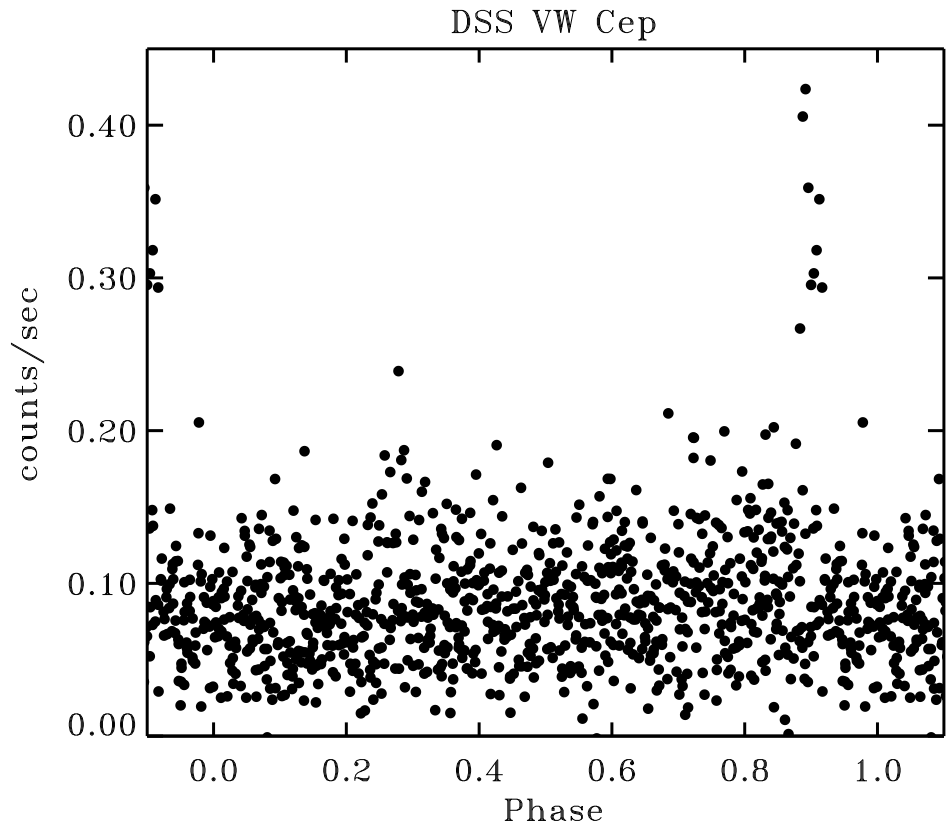,height=4.0in}}
\vskip 0.5in
\caption{\label{fig4}
The phase-folded light curve for VW~Cep.}
\end{figure}

\begin{figure}      
\centerline{\psfig{figure=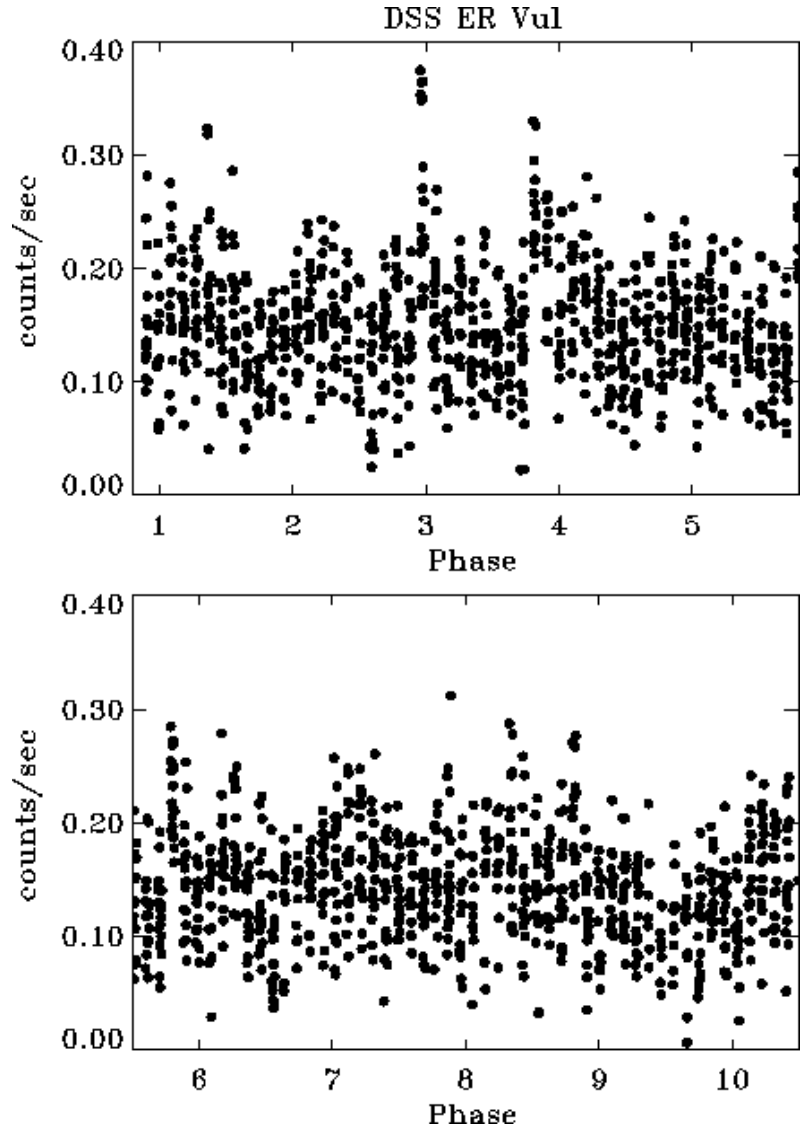,height=6.0in}}
\vskip 0.5in
\caption{\label{fig5}
The DSS light curve for ER~Vul. The system showed some low-level activity
most of the time.}
\end{figure}

\begin{figure}      
\centerline{\psfig{figure=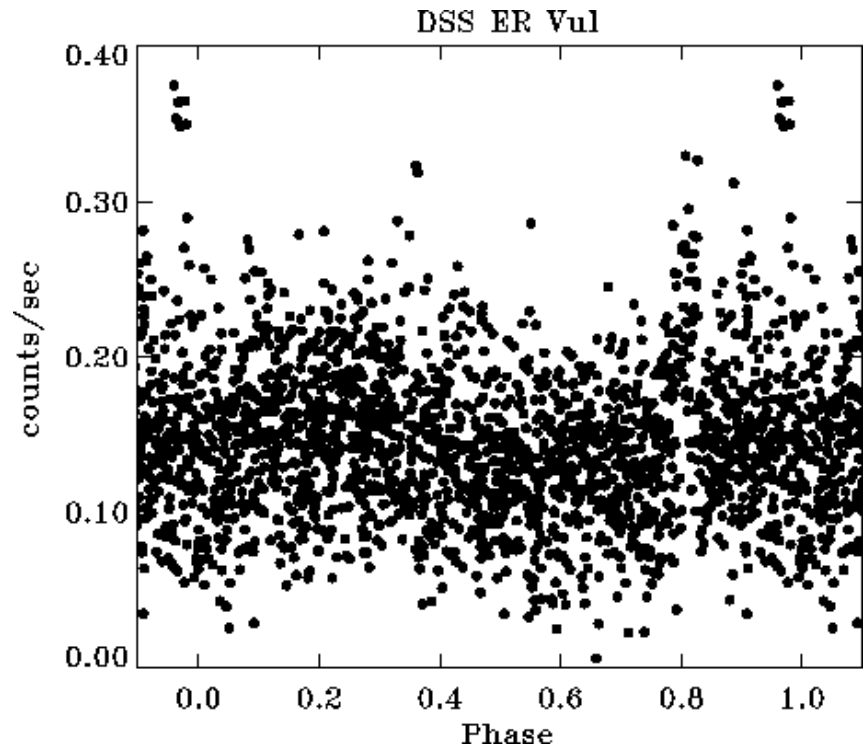,height=4.0in}}
\vskip 0.5in
\caption{\label{fig6}
The phase-folded light curve for ER~Vul. Note that the baseline level
was on the average highest at phases close to 0.2 and lowest close to
0.7, while most low-level outbursts took place in 
the phase interval 0.8 to 1.0.}
\end{figure}

\begin{description}

\item [44i~Boo~B:] The light curves are in  
Figures~\ref{fig1} and \ref{fig2}. The observed 
count-rate mean and median for 44i Boo were 0.340 
and 0.331 counts/sec. Some small activity 
reaching up to 0.6 counts/sec seemed to be present all 
the time. Only one moderate brightening at phase 
about  15.5 was observed. The data plotted in one 
orbital-phase interval (Fig.~\ref{fig2}) show a weak 
orbital-phase dependence with the lowest count rates 
at the primary (deeper) eclipse. This eclipse corresponds 
to occultation of  the smaller, slightly hotter 
component. However, a tendency of the data points 
to show some grouping in Fig.\ref{fig2} is spurious; it 
is due to commensurability of  stellar and satellite orbits. 

\item [VW~Cep:] The light curves are in 
Figures~\ref{fig3} and \ref{fig4}. The mean and 
the median count-rates were 0.088 and 0.083 
counts/second. One short flare reaching 
levels some five times the quiescent level was observed 
just before the primary eclipse at the phase 
of about 16.9 (Figure~\ref{fig3}). The variability of VW~Cep 
showed no phase relation (Figure~\ref{fig4}).

\item [ER~Vul:] The light curves are in 
Figures~\ref{fig5} and \ref{fig6}. The mean and the 
median count rates were 0.146 and 0.143 
counts/second. Several short increases in the 
DSS count rates by about 2 -- 3 times the average rate 
were observed during the run. 
They seemed to be more frequent in the orbital phases interval 
around 0.8 to 1.0, i.e. just before the deeper 
minimum (Figure~\ref{fig6}). 
The quiescent level seemed to show a low-amplitude 
sinusoidal modulation with a maximum at the 
orbital phases around  0.2 to 0.3. 
\end{description}

\begin{figure}      
\centerline{\psfig{figure=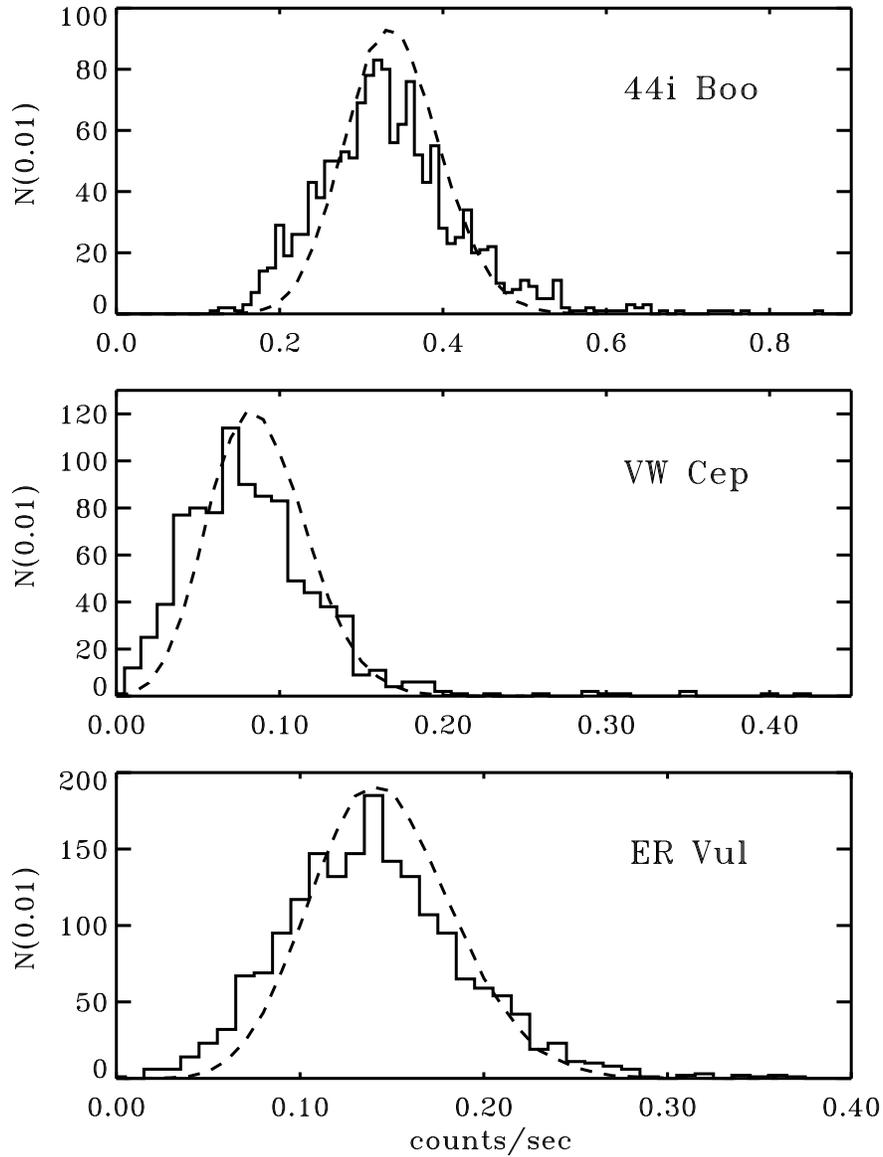,height=6.0in}}
\vskip 0.5in
\caption{\label{fig7}
Histograms of the DSS count rates for the three program stars. The
broken lines give the expected Poissonian distributions evaluated from
the mean count 
rates. The deviations of the observed histograms from the
Poisson distributions come from two sources: At the high
count rates, they were caused by genuine
increases of signal level; at the low count rates the Poisson
statistics was corrupted by inclusion of data from
shorter time bins (down to 50 sec).} 
\end{figure}

The DSS data for all three stars are shown 
in Figure~\ref{fig7} as histograms of the DSS count rates. 
The broken lines give the expected 
Poisson distributions for 100 second intervals based
on the mean count rates. Deviations from 
the Poisson distributions on the side of 
low counts are due to inclusion of shorter intervals, down to 
50 seconds, which led to an increased random 
noise. The deviations on the high side are due to 
flaring activity which, as we can see,  
was moderate in all of the three cases.

Since the DSS data indicated moderate 
orbital and flaring variability of any of the program stars, 
the spectral data (Section~\ref{spectra}) were analyzed 
assuming temporal constancy of the signal. However, 
some tests were made to see if any variability 
was visible in the strongest spectral lines or in 
combined spectra binned in four large phase 
intervals (two eclipses and two maxima). No 
variability which would be statistically 
significant was detected, mostly because of the low photon 
rates. 

\section{NEUTRAL HYDROGEN ABSORPTION}
\label{hydrogen}

Estimates of the interstellar hydrogen 
absorption are needed for interpretation of the spectral data. 
They have been made using the Center for EUV 
Astrophysics Internet facility called ``ISM 
Hydrogen Column Density Search Tool'' 
 where the database of previous determinations permits 
their evaluation on the basis of 
angular and spatial distances from the target object. The database is 
based on the compilation of Fruscione et al. 
\markcite{frusc94}(1994) with additions from Diplas \& Savage 
\markcite{ds94}(1994). To estimate the monochromatic 
absorption at a given distance, another 
Internet facility, called ``ISM Transmission Tool'' 
was used, this one being based on the study of 
Rumph et al. \markcite{rum94}(1994). 
The assumptions on the relative number of species were: 
N(He~I)/N(H~I) = 0.1 and N(He~II)/N(H~I) = 0.01.

\begin{description}
\item[44i~Boo~B:] The column density estimated for most 
of the neighbor stars is $\log N_H({\rm cm}^2) < 17.4$. 
This is consistent with the estimates 
based on X-ray and UV observations by 
Vilhu \& Heise \markcite{vh86)}(1986) and 
Vilhu et al. \markcite {vil88}(1988) 
of  $\log N_H({\rm cm}^2) < 18$. 
Dupree et al.\markcite{1996b} used  
$\log N_H({\rm cm}^2) = 18.0$, 
but this value seems to be too high. 
We have assumed $\log N_H({\rm cm}^2) = 17.5$ which is 
consistent with the visibility of emission lines in 
MW and LW spectra beyond 350 \AA.

\item [VW~Cep:] The database gives very widely 
scattered determinations of $N_H$. For a distant 
star $\kappa$~Cep (B9III), a few degrees away, 
$\log N_H({\rm cm}^2) < 18.1$. Vilhu \& Heise 
\markcite{vh86)} (1986) give 
$\log N_H({\rm cm}^2) < 18$. Assuming that the spectrum of 
VW~Cep has similar {\it relative\/} 
intensities of lines as in 44i~Boo~B for which we assumed
$\log N_H({\rm cm}^2) = 17.5$ (Section~\ref{spectra}),
 we obtained $\log N_H({\rm cm}^2) \simeq 18.0$ for VW~Cep, 
the value used hereinafter.

\item [ER~Vul:] Because of the larger distance than 
to the two contact binaries, one would expect 
stronger neutral hydrogen absorption. 
The database suggests values at the level of 
$\log N_H({\rm cm}^2) \le 19$. 
The X-ray estimate by White et al. \markcite{white87}(1987) of 
$N_H({\rm cm}^2) = 6_{-5}^{+14} \times 10^{18}$ 
permits a very wide range of possibilities, 
$18 < \log N_H({\rm cm}^2) < 19.3$. 
The more risky (than for VW~Cep) 
assumption that the overall {\it relative\/} 
line intensities in 44i~Boo~B and ER~Vul are 
similar (Section~\ref{spectra}) leads to an estimate $\log 
N_H({\rm cm}^2) \simeq 18.5$. For such 
absorption the emission lines in the MW and LW 
regions would not be visible, as is actually the case.
\end{description}

\section{EUVE SPECTRA}
\label{spectra}

\subsection{General properties}

\begin{figure}      
\centerline{\psfig{figure=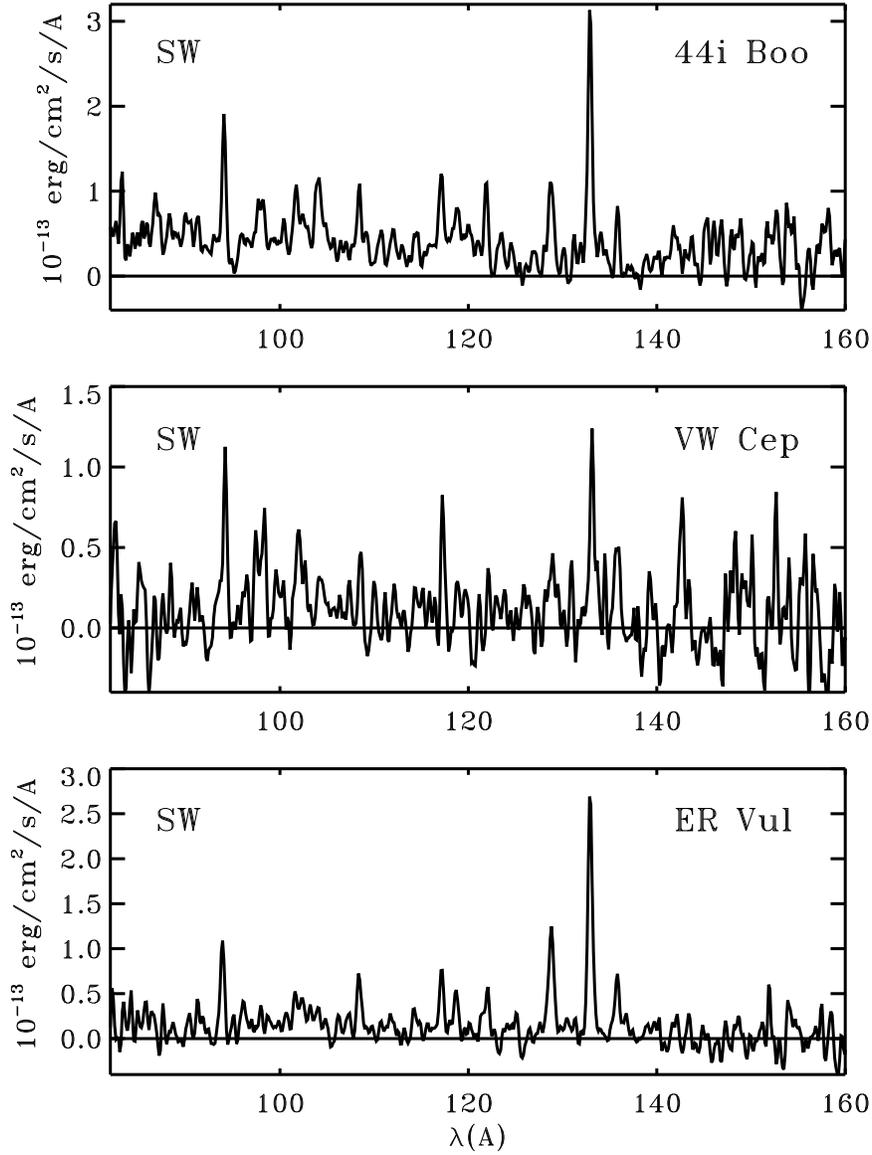,height=6.0in}}
\vskip 0.5in
\caption{\label{fig8}
The integrated SW spectra of the three program stars expressed in
monochromatic fluxes. The units on the  vertical axis are 
$10^{-13}$ erg/cm$^2$/s/\AA.}
\end{figure}

\begin{figure}      
\centerline{\psfig{figure=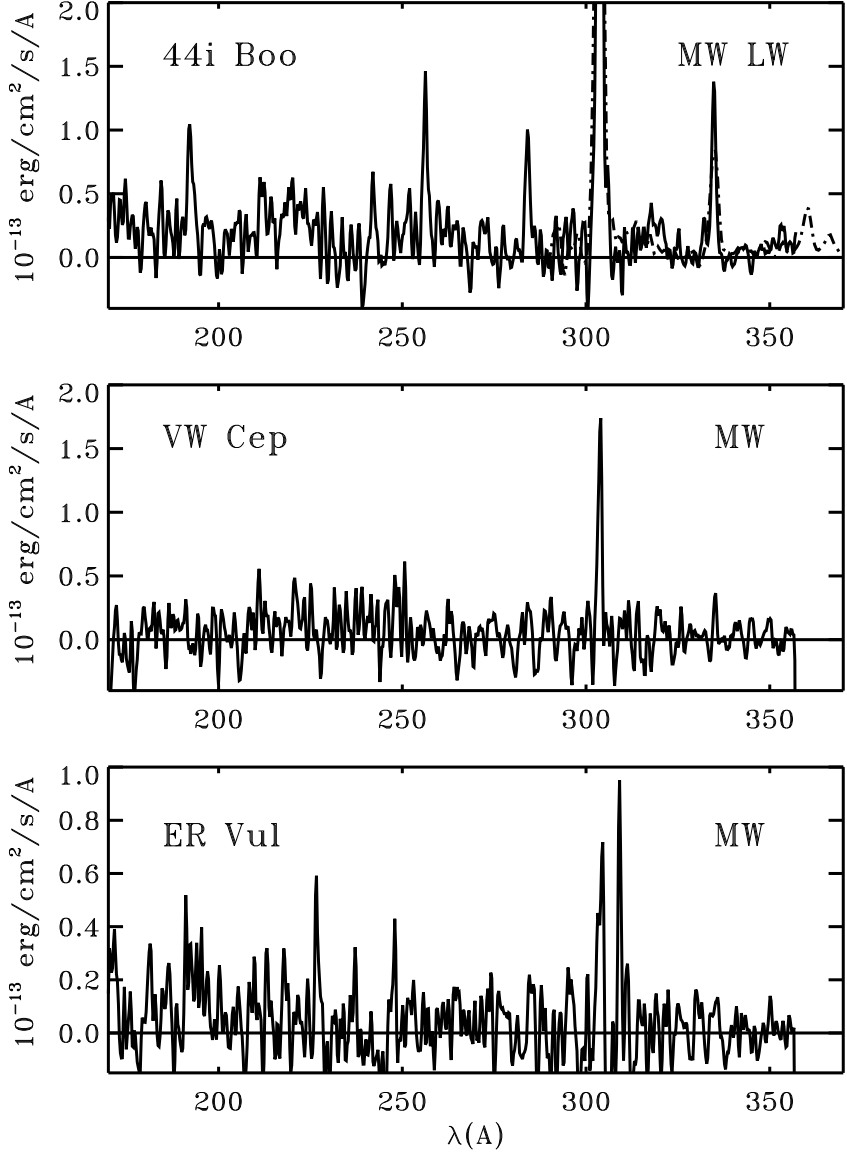,height=6.0in}}
\vskip 0.5in
\caption{\label{fig9}
The integrated MW spectra of the three program shown in the same
format as in Fig.~\ref{fig8}. The LW spectrum
is over-plotted in the case of 44i~Boo (broken line); 
the other two stars do not show
any stellar features in the LW band. Note that for ER~Vul, the He~II
line is corrupted by the large Poissonian fluctuations entering through
subtraction of the strong background geocoronal feature.}
\end{figure}

The observed spectra are shown in Figures~\ref{fig8} and 
\ref{fig9}. They are plotted as observed fluxes, 
uncorrected for neutral-hydrogen absorption. 
The spectra have been Gaussian smoothed with a filter 
having $FWHM = 4.7$ pixels for each spectral band 
(0.0675, 0.135 and 0.270 \AA). Thus, they 
are still slightly over-sampled as the actual resolution 
was close to 7 original pixels in each of the bands. 
The spectra show the 
strong effects of the interstellar absorption for 
VW~Cep and especially for ER~Vul, where even 
the MW band (170 -- 350 \AA) is practically 
devoid of emission lines, in spite of the presence of 
strong lines in the SW band. The usually very strong 
He~II 304 \AA\  line in ER~Vul is absorbed 
to such a degree that the large Poissonian 
noise entirely dominates the shape of the line. Only 
44i~Boo~B shows moderate absorption in the MW band 
and in the short wavelength part of the LW band. The He~II 304 \AA\  
line is very strong in 44i~Boo~B and extends beyond the displayed
range of the fluxes in Figure~\ref{fig9} with 
$f_\lambda = 10^{-12}$ erg/cm$^2$/s/\AA. 
The integrated emission line fluxes are 
listed in Table~\ref{tab2}. They are given only for lines for 
which the formal errors, resulting from combined 
uncertainties in the line flux as well as in the 
subtracted  background, were smaller than 30\%; 
usually, the errors were at the level of about 10\% to 15\%. 

Because hydrogen absorption is moderate in the SW band 
(80 -- 160 \AA) even for relatively large neutral-hydrogen column
densities, one can compare 
directly the spectra in the three studied stars in this spectral
region. They turn out quite similar in terms of {\it relative\/} 
strengths of the emission lines. This can be used as a check on  
our estimates of  hydrogen absorption assuming that this similarity
extends into the MW and LW regions. The 
effects of absorption are moderate (factor of less than 3-times) 
over the wide band between 80 -- 450 \AA\  for 
$\log N_H({\rm cm}^2) < 18$, but become rapidly 
more severe for higher column densities and longer wavelengths.
The estimates of $\log N_H({\rm cm}^2)$ were
based on the relative strength of lines in 44i~Boo~B 
and the remaining two
stars assuming that for the former star, the absorption 
is $\log N_H({\rm cm}^2) = 17.5$. The 
results were 18.0 for VW~Cep and 18.5 for ER~Vul, with
uncertainties at the level of 0.3 to 0.5 in $\log N_H$. 

\subsection{Emission measure solution for 44i~Boo~B}

Our spectra of VW~Cep and ER~Vul contain too 
few lines, observed with a too low signal-to-noise 
ratio to contemplate emission measure determinations. 
The additional uncertainty is due to the poorly known, 
but probably large, neutral hydrogen absorption, especially 
for ER~Vul. Only for 44i~Boo~B the data are of 
sufficient quality to obtain a differential emission 
measure distribution. The procedure described by 
Schmitt et al. \markcite{sch96}(1996) was 
followed very closely in that only emission lines 
of various iron ions were used. This way, the large 
uncertainties related to unknown relative 
abundances of  coronal species become less important. 
The integral equations for observed flux in 
line $j$ of the shape: $\phi_j = {1 \over {4 \pi d^2}} 
\int D(T) \,G_j(T)\, d\,\log T$ were approximated 
by discrete representations: 
$\phi_j = {1 \over {4 \pi d^2}} \Sigma \, D(T_i)\, G_{i,j}\,
\Delta \log T$. 
The over-determined system of equations was 
solved using the SVD technique. The solutions 
were obtained by adding successively more singular 
values and considering all non-negative DEM distributions with
progressively better fit to the data.

We used all 174 lines of iron ions in ionization stages 
Fe~IX to Fe~XXIV, as tabulated by 
Brickhouse, Raymond \& Smith \markcite{brick95}(1995). 
These lines cover formation 
temperatures within $5.4 < \log T < 7.8$. 
It should be stressed that we used all the tabulated lines, 
although the very small subset of only 16 lines
had the observed values. The rationale is that the lines 
which are absent or weak are equally 
important as those that are observed. In fact, experiments 
have shown that solutions based only on 
visible lines are not sufficiently constrained to the point of 
becoming basically meaningless. Various 
values of the temperature increment were tried and a 
relatively large value $\Delta T = 0.3$ 
(within $5.7 < \log T < 7.5$) was finally adopted. This gave 
only 7 discrete values of DEM. We found that 
solutions with finer sampling or extending too far in 
the temperature scale would lead to negative 
values of DEM even for relatively low order of singular values. 

\begin{figure}      
\centerline{\psfig{figure=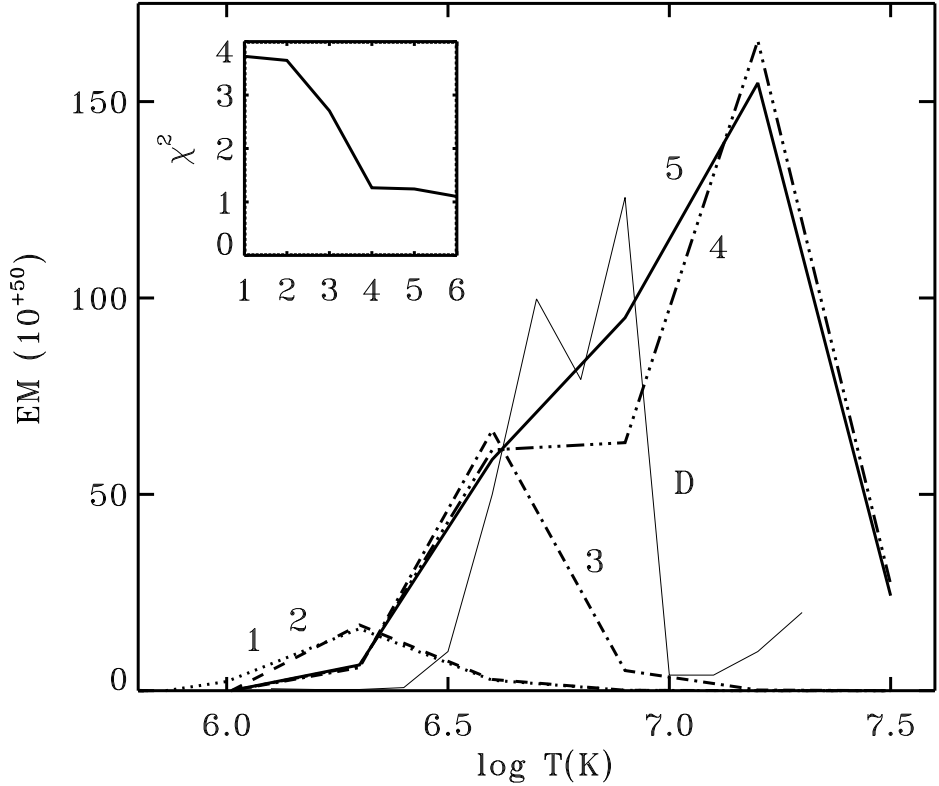,height=4.0in}}
\vskip 0.5in
\caption{\label{fig10}
The differential emission measure for 44i~Boo~B evaluated using the
approach of successive addition of singular-value solutions, as described
in the text. The DEM was derived for 7 independent
temperature bins, each $\Delta \log T = 0.3$ wide, without any
constraint on the smoothness,  but with a requirement of
non-negativity, as in Schmitt et al. 
(1996). Although none of the solutions up to the 5th order would be
negative anywhere, they showed no improvement in
$\chi^2$ beyond the 4th order. The
successive orders are marked by appropriate numbers in the figure and
the drop in the $\chi^2$ measure of the fit is shown in the insert. The
solution of Dupree et al. 
(1996) is marked by letter ``D'' and is shown by the thin line in the
figure.} 
\end{figure}

The DEM solution is shown in Figure~\ref{fig10}. Although 
the solution looks plausible, we are not 
satisfied with it for various reasons. 
The main problem is its uniqueness. The prescription of by 
Schmitt et al. \markcite{sch96}(1996) invokes 
buildup of  successive approximations of the DEM 
by inclusion successively more singular values,
as long as the resulting function is everywhere 
positive.  In our case, it was found that 
the first two approximations to the DEM function gave a 
single peak between $6.0 < \log T < 6.5$. 
The 3rd solution migrated to higher temperatures with the 
peak at $\log T \simeq 6.5$, 
and the 4th and 5th solutions migrated further to give a well defined 
peak at $\log T \simeq 7.2$. All lower 
order solutions are confined within the 5th order solution.
From the changes in $\chi^2$, we decided to 
stop at this point. Also, the still higher-order solutions start 
giving slightly negative values at 
$\log T \simeq 5.7 - 6.0$, with no improvement in $\chi^2$.
From the SVD solution and from further experiments 
with the DEM to spectrum mapping, we got 
a general impression that the DEM changes 
quickly at the high-temperature end for relatively small 
changes in the strength of the ``hot'' lines of 
Fe~XXIII and Fe~XXIV at 132 \AA\  and 192 \AA. 

We note that the new solution is different 
from the solution of Dupree et al. \markcite{dup96b} 
(1996) which showed a relatively narrow peak at 
$\log T \simeq 6.75$, but {\it  both solutions are
based on the same observations\/}.
Use of the same hydrogen absorption as that 
assumed by Dupree et al. of $\log N_H = 18$ 
(which we feel was too large) does not remove 
the discrepancy.  The solution of Dupree et al. 
was apparently made by trial and error. 
We feel that the present solution, which was done in a more objective
way, directly shows the difficulty of 
deriving the DEM from moderate-quality data as 
those available for 44i~Boo~B. In particular, the 
migration of solutions from low-order 
singular-value solutions concentrated at relatively low 
temperatures to higher-order solutions 
encompassing low-order ones, but peaking at higher 
temperatures directly shows the poor constraining of the problem. 

\section{COMPARISON OF LOWER CORONA ACTIVITY}
\label{comparison}

The EUVE spectra of 44i~Boo~B, VW~Cep, 
and ER~Vul offer a possibility to compare  
activity levels of coronal emissions 
for stars with very short rotation periods below one day. 
Similar comparisons 
were frequently made on the basis of IUE and 
X-ray emissions, culminating in the well-known 
``saturated-activity'' paper of  
Vilhu \& Walter \markcite{vw87}(1987). 
Here we will augment the new data by the observations of AB~Dor, a
single rapidly-rotating young star (Rucinski et al.
\markcite{ruc95} 1995). 
Its period rotation period of 0.51 day falls approximately half way
between the periods of 44i~Boo~B and VW~Cep on one hand and ER~Vul on
the other hand. All  stars have rotation periods shorter than one
day. Of crucial importance is the fact that this is the
only single star in the regime of very short periods accessible to
EUVE observations. We do not use here the old X-ray observations as
they were made with low spectral resolution.

The comparison is made using representative, strong emission
features. We use here 
the Mg~II 2800 \AA\  feature for chromospheric 
plasma at $\log T \simeq 4.2$, the C~IV 1550 \AA\  
feature for the transition-region plasma at 
$\log T \simeq 5.1$, both observed with the IUE, 
adding the new EUVE results for the 
He~II 304~\AA\  line (also formed at about
$\log T \simeq 5.1$) and for the strong Fe~XXIII 132~\AA\ 
line which is formed at $\log T \simeq 7.1$. 
Unfortunately, the gap between $\log T \simeq 5.1$ 
and $\log T \simeq 7.1$ is very wide: 
the lines forming at the intermediate temperatures 
$6.1 < \log T < 6.4$ and visible in EUVE 
spectra of nearby objects are too heavily absorbed in VW~Cep, 
ER~Vul and AB~Dor for reasonable comparison. 
Only the Fe~XVI 335 \AA\ line ($\log T \simeq 6.4$) 
is visible in all four stars, 
but its strength is very uncertain due to large neutral-hydrogen 
corrections. The IUE observations were taken 
from the original sources, frequently cited later in 
various summary papers: 
Rucinski \& Vilhu \markcite{rv83}(1983), Vilhu \& Rucinski  
\markcite{vr83}(1983), Rucinski \markcite{ruc85c} 
(1985c) and Vilhu \& Heise \markcite {vh86} (1986).

\begin{figure}      
\centerline{\psfig{figure=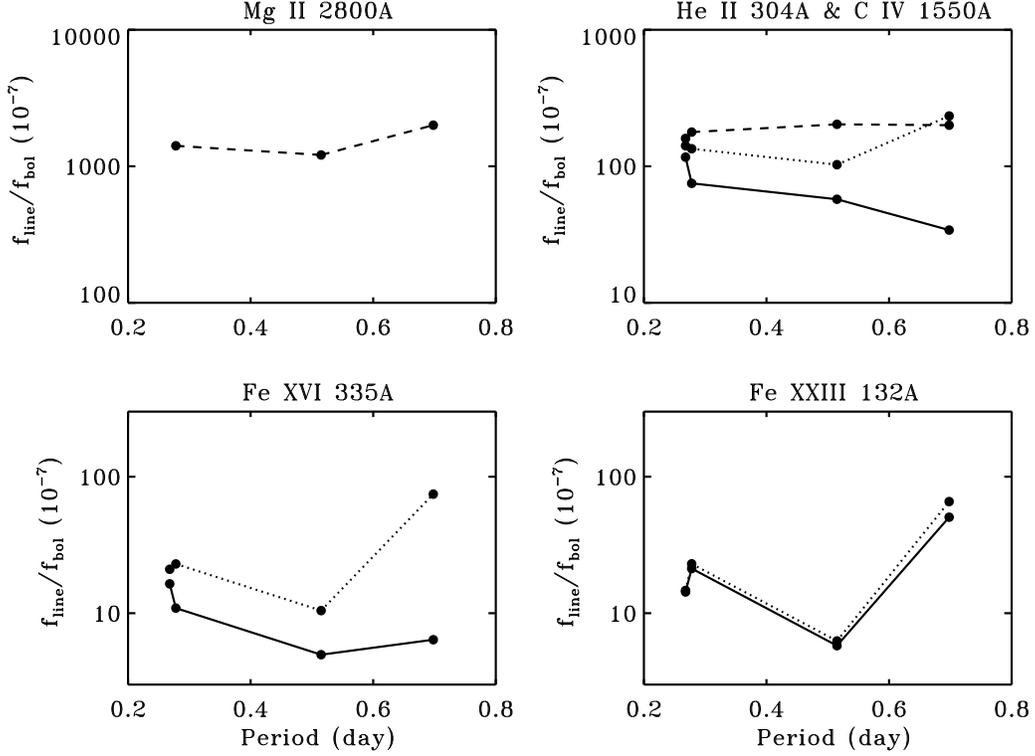,height=4.0in,angle=90}}
\vskip 0.5in
\caption{\label{fig11}
Comparison of the relative emission-line fluxes,
$f_{line}/f_{bol}$, for the three program stars and for AB~Dor, a
single rapidly-rotating (0.51 day) K-type dwarf. In each panel,
the two contact binaries are on the left, ER~Vul is on the right, and
AB~Dor is in the middle.
The panels give the relative fluxes for four typical temperatures of
line formation: $\log T \simeq 4$ (Mg~II 2800~\AA),
$\log T \simeq 5.0 - 5.1$ (He~II 304~\AA\ and C~IV 1550~\AA),
$\log T \simeq 6.4$ (Fe~XVI 335~\AA) and $\log T \simeq 7.1$
(Fe~XXIII 132~\AA). The IUE data are connected 
by broken lines whereas the
EUVE data are connected by continuous (observed) and dotted
(H~I-corrected) lines. The formal
measurement errors are not shown as they
are usually very small for the strong lines shown here; in fact,
systematic errors, which are very difficult to estimate, dominate in
determining the uncertainties of the data.}
\end{figure}

Figure~\ref{fig11} contains comparison of the four 
stars in terms of the relative emission-line fluxes, 
$f_{line}/f_{bol}$. Arguably, this is the only 
meaningful quantity which can be used for comparison of  
contact (44i~Boo~B, VW~Cep) and detached (ER~Vul) 
binary stars with a single (AB~Dor) star, 
as the major differences in the radiating 
areas are then automatically taken into account. Figure~\ref{fig11}
shows that all four stars are 
practically identical in their chromospheric (Mg~II) and 
transition-region (He~II and C~IV) emission 
properties, but relatively large differences are 
observed in high-temperature emissions. 
Corrections for the neutral hydrogen absorption are large 
for the transition-region He~II 304 \AA\ line, 
especially for ER~Vul, but we seem to see the same 
lack of period dependence as observed in 
the C~IV line, which is consistent with the ``saturated'' regime of
activity in these stars. The absorption corrections are even larger 
for the Fe~XVI 335 \AA\  line. In this line, 
AB~Dor seems to be the least active among the
four stars. The largest deviation of AB~Dor relative to the three
binaries is observed for the ``hot'' line of Fe~XXIII at 132 \AA.
Obviously, without an extensive temporal coverage, 
it is impossible to tell if 
AB~Dor was observed in a minimum of its 
dynamo cycle or is always less active than the 
bracketing (in the period domain) binary stars. 
We note, however, that Drake \markcite{drake96} 
(1996) already pointed out that the chromospheric 
parts of the emission-measure distributions are 
similar (or predictable)  in most active stars, 
whereas large differences between stars in the
emission-measure distributions appear at high temperatures, for
$\log T > 10^6$.

\section{CONCLUSIONS}
\label{conclusions}
Three very close binary stars with the orbital periods shorter than
one day, 44i~Boo~B, VW~Cep and ER~Vul, show very similar levels of
EUV emission line fluxes. The two former are contact binaries with a
period close to quarter of a day, the latter is a very close, but
detached binary with a the period almost three times longer. During
the observations, which consisted of separate, several days-long,
continuous pointings of the EUVE satellite, none of the stars showed 
large flares or prolonged
brightenings; however, the DSS photometry data
indicate that low level activity was almost constantly present in each
case. These low-level variations were disregarded when forming
time-integrated EUV spectra.

The spectra indicate relatively strong neutral hydrogen absorption in
the case of VW~Cep and especially ER~Vul. Thus, it was impossible to
attempt differential emission measure determinations for these
stars. Such a determination was possible for 44i~Boo~B, but the
quality is moderate (at best) because of the small number of emission
lines with well measured fluxes. The solution followed closely the
methodology of Schmitt et al.\markcite{sch96}(1996) 
in that (1)~to avoid problems
with uncertain relative abundances, only iron lines were
used, and (2)~the technique of singular-value decomposition was used
to prevent over-interpretation of the information content in the
spectrum. The result on the differential emission measure (DEM) for
44i~Boo~B shows lack of any obvious preference of a 
specific interval of the temperature, with the DEM rising within the
accessible interval $6.0 < \log T < 7.2$. The determination strongly
depends on inclusion of individual lines so that this determinations
of the DEM must be still considered as preliminary.

Strengths of a few strong, well observed emission lines were compared
for the three binary stars and for AB~Dor, the only single,
rapidly-rotating {\it single\/} star with the rotation period shorter
than one day which is accessible to the EUVE observations
(Rucinski et al.\markcite{abdor} 1995). While the emission feature
strengths, expressed in $f_{line}/f_{bol}$, for lines
forming at temperatures below $\log T \simeq 5$ show the expected
uniformity for all four stars (possibly due to the ``saturation''),
the hotter lines were observed to be definitely weaker in AB~Dor than
in the three binaries. It is unclear whether this was due to some
temporal variability or is a permanent feature which distinguishes
single stars from binaries in this very extreme range of stellar
rotation conditions. 

\acknowledgments
Thanks are due to Jeneen Sommers for her help during the reductions of
the EUVE data at the Center for EUV Astrophysics.

\newpage
 
\begin{deluxetable}{lccc}     
\tablecaption{Stellar and observation data \label{tab1}}
\tablewidth{0pt}
\tablehead{
\colhead{Parameter} & \colhead{44i~Boo~B} & \colhead{VW~Cep} 
& \colhead{ER~Vul}
}
\startdata
    & HD~133640 & HD~197433 & HD~200391 \nl
V   &   (5.9)   &   7.30    &  7.25     \nl
B-V &   (0.8)   &   0.85    &  0.59     \nl
Sp type & (G8:) & (K0 - K2:) & G1V + G1V \nl
Distance (pc) &  13 & 24    & 46        \nl
Adopted $\log N_H$ (cm$^{-2}$) & 17.5 & 18.0 & 18.5 \nl 
B.C. &  $-0.2$  &  $-0.24$  & $-0.06$   \nl
$f_{bol}$ (erg/cm$^2$/s) & $1.3 \times 10^{-7}$ & $3.7 \times 10^{-8}$ & 
$3.3 \times 10^{-8}$ \nl
Adopted $P$ (day) & 0.26781856 & 0.2783076 & 0.698095 \nl
Adopted $T_0$ (JD) & 2449489.483 & 2448862.521 & 2446328.984 \nl
EUVE source & J1503+47.6 & J2037+75.5 & J2102+27.8 \nl
EUVE project & 93-013 & 93-013 & 94-011 \nl
Start UT (geo) & 2 May 1994 9:38 & 30 Jan 1995 1:29 &
20 Sep 1995 3:25 \nl
Stop UT (geo) & 7 May 1994 18:37 & 4 Feb 1995 2:08 &
27 Sep 1995 9:26 \nl
Start JD (hel) & 2449474.904 & 2449747.562 & 2449980.646 \nl
Stop JD (hel) & 2449480.278 & 2449752.589 & 2449987.897 \nl
Nominal exp (sec)   & 130,000 & 100,000 & 200,000 \nl
Actual exp SW (sec) & 135,462 & 88,951 & 181,394 \nl
Actual exp MW (sec) & 132,051 & 88,585 & 179,510 \nl
Actual exp LW (sec) & 133,615 & 88,027 & 181,282 \nl
\enddata
\tablecomments{See the text for full explanations.}
\end{deluxetable}

\begin{deluxetable}{lcrrrrrrrr}      
\tablecaption{Observed EUV emission line fluxes \label{tab2}}
\tablewidth{0pt}
\tablehead{
\colhead{Iron} & \colhead{band} & 
\multicolumn{2}{c}{44i~Boo~B} &
\multicolumn{2}{c}{VW~Cep}    &
\multicolumn{2}{c}{ER~Vul}    &
\multicolumn{2}{c}{AB~Dor}    \\
\colhead{~ion} & &
$\lambda$(\AA) & $\phi$ &
$\lambda$(\AA) & $\phi$ &
$\lambda$(\AA) & $\phi$ &
$\lambda$(\AA) & $\phi$
}
\startdata
11? & SW & 86.9 & 82 &\nodata &\nodata &\nodata &\nodata &\nodata &\nodata \nl
18  & SW & 94.1 & 109 & 94.2 & 55 & 93.9 & 62 & 94.2 & 22 \nl
21  & SW & 98.0 &  98 & 98.3 & 41 &\nodata &\nodata & 98.2 & 13 \nl
19  & SW &\nodata &\nodata &\nodata &\nodata & 101.6 & 24&\nodata &\nodata \nl
21  & SW & 102.1& 138 & 102.1& 61 & 102.3 & 31& 102.4& 14 \nl
18  & SW & 104.1& 103 & 104.3& 27 & 104.0 & 23& 104.1& 10 \nl
19  & SW & 108.4& 64  & 108.5& 25 & 108.4 & 43& 108.5& 17 \nl
22 & SW  & 117.1 & 91 & 117.3 & 35 & 117.1 & 46 & 117.3 & 16 \nl
20 & SW & 118.7 & 76 &\nodata &\nodata & 118.7 & 31 & 118.7 & 10 \nl
19 & SW & 120.0 & 57 &\nodata &\nodata &\nodata &\nodata &\nodata &\nodata \nl
20 & SW & 121.9 & 65 & 122.2 & 12 & 121.9 & 35 & 122.0 & 10 \nl
21 & SW & 128.8 & 80 & 128.8 & 35 & 128.8 & 92 & 129.0 & 16 \nl
23 & SW & 132.9 & 186 & 133.2 & 79 & 133.0 & 167 & 133.0 & 35 \nl
22 & SW & 135.9 & 38 & 135.7 & 40 & 135.8 & 45 & 136.0 & 11 \nl
24 & MW & 192.5 & 206&\nodata &\nodata & 191.2 & 40 & 192.2 & 25 \nl
14 & MW & 211.9 & 122& 211.0 & 52 &\nodata &\nodata & 212.1 & 11 \nl
He II& MW& 256.4 & 223 &\nodata &\nodata &\nodata &\nodata & 256.3 & 23 \nl
He II& MW& 303.7 &1580 & 303.7&219&304.2& 89 &\nodata &\nodata \nl
He II& LW& 303.6 &1454 & 302.8&336&303.4&136 & 303.1 & 344 \nl
16 & MW & 334.8 & 193 & 335.1 & 33 & 335.4& 17 & 335.1 & 18 \nl
16 & LW & 334.9 & 234 & 333.3 & 48 & 335.4& 25 & 333.8 & 42 \nl
10 & LW & 353.4 & 42   &\nodata &\nodata &\nodata &\nodata & 353.8 & 9 \nl
16 & LW & 360.0 & 126 &\nodata &\nodata &\nodata &\nodata & 359.9 & 16 \nl
12 & LW & 366.2 & 58  &\nodata &\nodata &\nodata &\nodata &\nodata &\nodata \nl
15 & LW & 416.8 & 34  &\nodata &\nodata &\nodata &\nodata & 416.6 & 18 \nl
\enddata
\tablecomments{The emission line fluxes $\phi$ are
in $10^{-15}$ erg/cm$^2$/s. The empty boxes signify that either the
line was not detectable or the measurement carried a formal error larger
than 30\%.}
\end{deluxetable}

\end{document}